# *Ab Initio* Modeling Of Friction Reducing Agents Shows Quantum Mechanical Interactions Can Have Macroscopic Manifestation

J. D. Hernández Velázquez[1*], J. Barroso – Flores[2*], and A. Gama Goicochea[3†]

[1]Centro de Investigación en Ciencias Físico – Matemáticas, Facultad de Ciencias Físico - Matemáticas, Universidad Autónoma de Nuevo León, San Nicolás de los Garza 66450, Nuevo León, Mexico

[2]Centro Conjunto de Investigación en Química Sustentable UAEM-UNAM, Carretera Toluca-Atlacomulco Km 14.5, Unidad San Cayetano, Toluca 50200, Estado de México, Mexico

[3]División de Ingeniería Química y Bioquímica, Tecnológico de Estudios Superiores de Ecatepec, Av. Tecnológico s/n, Ecatepec 55210, Estado de México, Mexico

## ABSTRACT

Two of the most commonly encountered friction reducing agents used in plastic sheet production are the amides known as erucamide and behenamide, which despite being almost identical chemically, lead to markedly different values of the friction coefficient. To understand the origin of this contrasting behavior, in this work we model brushes made of these two types of linear – chain molecules using quantum mechanical numerical simulations under the Density Functional Theory at the B97D/6-31G(*d*,*p*) level of theory. Four chains of erucamide and behenamide were linked to a 2X10 zigzag graphene sheet and optimized both in vacuum and in continuous solvent using the SMD implicit solvation model. We find that erucamide chains tend to remain closer together through π – π stacking interactions arising from the double bonds located at C13–C14, a feature behenamide lacks and thus a more spread configuration is obtained with the latter. It is argued that this arrangement of the erucamide chains is responsible for the lower friction coefficient of erucamide brushes, compared with behenamide brushes, which is a macroscopic consequence of cooperative quantum mechanical interactions. While only quantum level interactions are modeled here, we show that behenamide chains are more spread out in the brush than erucamide chains as a consequence of those interactions. The spread out configuration allows more solvent particles to penetrate the brush, leading in turn to more friction, in agreement with macroscopic measurements and mesoscale simulations of the friction coefficient reported in the literature.

---

[*] Equally contributing authors.
[†] Corresponding author. Electronic mail: agama@alumni.stanford.edu . Telephone: +52 (55) 5000 2300.

Additives have been used for years as sliding agents to modify the tribological properties of thin films[1-3]. Functionalized polymethylmethacrylate (PMMA)[4] and polyolefines (POs)[3, 5-9] surfaces with brushes composed by fatty acids (FAs) have proved to be useful to reduce the coefficient of friction (COF) of these materials. Even more, experimental results have shown that the COF depends on the hydrocarbon chain length of the FAs[4,10], the extrusion time[11], the surface concentration of FAs[8,9,12], and the aging of the sample[5,9]. One of the most popular slip agents is erucamide; this fluid-friction reducing molecule is obtained from naturally occurring fats and oils frequently used in the polymer industry[6-9,13]. However, there are other FAs used to reduce the COF, such as behenamide[4,7], where the only structural difference between these additives is the existence of a double bond between the C-13 and C-14 atoms in the erucamide molecule (see Fig. 1). Although these molecules are very similar, experiments have shown that their roles as slip additives are certainly different. For example, it has been found that the COF of PMMA sheets with erucamide concentration of 0.05% is ~ 0.09, while with behenamide at a concentration of 0.10% it is ~ 0.19 under the same normal load, sliding speed and at a temperature of 25°C [4]. Also, experiments with low – density polyethylene films of very similar bulk loadings of erucamide and behenamide (1010 ppm and 1080 ppm, respectively), exhibit COF for films with behenamide of ~ 0.48, while for films with erucamide the COF is only ~ 0.18 [7].

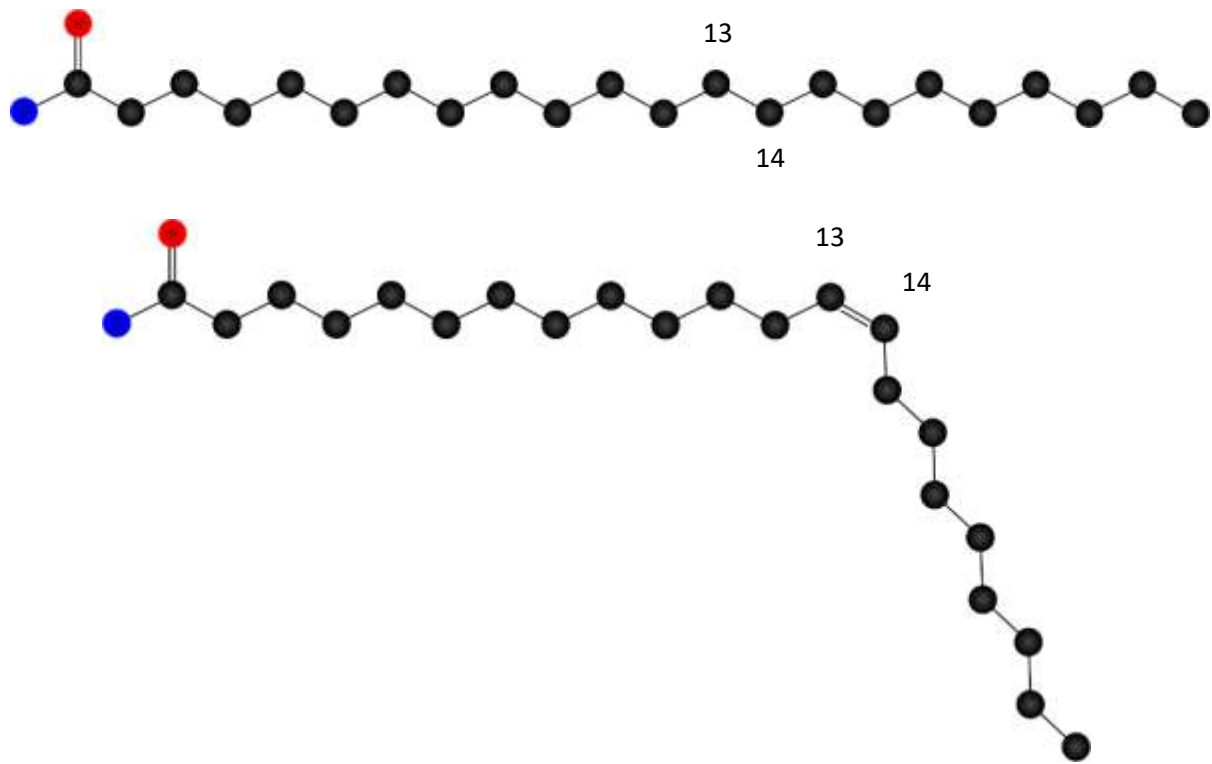

**Figure 1**. Chemical structures for behenamide (top) and *cis*-erucamide (bottom). The double bond between the C 13 and C14 atoms is indicated with the numbers.

It is well known that the COF of surfaces covered with FAs decreases with increasing the chain length[10,14]. Experimental data show that at fixed sliding velocity, temperature and load, the COF decreases with increasing grafting density of the FAs[5-7,9,12]. On the modeling side, molecular dynamics (MD) has turned out to be a very useful tool, capable of reproducing the trends found in experiments, albeit in some instances the values of the COF are slightly different from those obtained in experiments, because the timescales usually differ between those two approaches[16-18]. MD studies shed light on the influence of unsaturated hydrocarbon chains upon the performance of FA as slip agents, and some studies[16,18] suggest that the presence of a double bond in the fatty acid chains prevents the penetration of the lubricant into the brush, causing less layering near the surface on which the brush is grafted. This behavior correlates with the COF as follows: for unsaturated chains, the COF depends weakly on the surface grafting density, while for saturated chains the COF increases with decreasing grafting density[18]. Additionally, when the sliding velocity is varied, the COF is found to increase with sliding velocity for saturated chains but not for unsaturated chains[15]. Recent mesoscopic scale simulations of surfaces covered with erucamide brushes[19] have been able to predict the trends and the quantitative values of the COF at fixed sliding velocity as well, for increasing grafting density. However, by their very nature, mesoscopic or coarse-grained simulations integrate out internal degrees of freedom of molecules, such as the double bond between C-13 and C-14 atoms in the erucamide molecule, which is the only structural difference between it and the behenamide molecule. Yet, this apparently small difference turns out to lead to markedly different values of the macroscopically measured COF for surfaces under flow covered by brushes made up of those two types of slip agents.

Surface properties such as the COF of materials functionalized with FAs, are found to depend principally on the surface coverage, and researchers report that the COF decreases as the surface coverage increases[7, 9, 14]. It is noteworthy that after a critical surface coverage there appears a plateau in the value of the COF, which does not change if the surface coverage[14] or the load[7] are increased. It is also important to note that the surface coverage depends on the load, as follows[7, 9]. A high load leads to high surface coverage; on the other hand, the surface coverage increases under a given load with the aging time of the sample as the extrusion proceeds. However, there is a critical surface coverage such that at larger values of it, the COF reduction is minimal[5, 9]. Additionally, the amides migration to the surface increases with their concentration in the plastic matrix[6], which leads to the reduction of the COF. There are however oscillations in the COF at small surface coverage because the coat is in the "mushroom" regime and after a critical value the plateau arises, signaling the transition from mushroom to the brush regimes[14].

In reference[19] the COF for erucamide chains was predicted in two different situations, one where the chains were all grafted to the surface, and the other where some chains were freely dispersed in the fluid confined by two parallel sheets, while the rest were grafted to the surfaces. The results showed that the COF reaches a plateau in both situations once the surface coverage has reached certain value. However, qualitative differences do appear at low coverage, where the COF oscillates when there are only grafted chains, while it decreases monotonically when there are additionally free chains[19]. These features are only important at low coverage because the chains are in the so – called mushroom regime and do not yet cover uniformly the surfaces. Lastly, in this work we present arguments as to why erucamide and behenamide chains at the same surface coverage lead to different values of the COF.

Better understanding of the origin of the contrasting tribology between erucamide and behenamide must be sought from a quantum mechanical viewpoint. In this work, we attempt to understand through *ab initio* quantum mechanical numerical simulation how the existence of the double bond in the hydrocarbon chains of FAs, particularly in the C-22 erucamide molecule, is responsible for its qualitatively and quantitatively different collective behavior from that of the behenamide. With that purpose in mind, we have modeled the erucamide and behenamide brushes with four molecules of each type, grafted on a graphene surface, and optimized their structure, as detailed in the following paragraphs.

All calculations were performed with the Gaussian 09 suite of programs[20]. Four erucamide molecules were bound to a 2X10 zigzag graphene sheet and were later optimized under the Density Functional Theory (DFT), namely at the B97D/6-31G(*d,p*) level of theory using Grimme's dispersion – corrected functional[21]. Both hydrocarbon chains under study, behenamide and erucamide, respectively, were placed four bonds apart from each other on the graphene sheet so as to have a maximum density without overcrowding their van der Waals surface, which leads to a 4.88 Å separation as measured between neighboring nitrogen atoms. This high density scheme is selected to investigate the limiting case for which electronic interactions between hydrocarbon chains are maximized without invoking a second surface. All carbon atoms in the graphene sheet were frozen to provide a rigid scaffold for the free interaction of the hydrocarbon chains under study as well as to decrease the computation time. A frequency analysis was performed on the Hessian matrix to confirm the presence of local minima at the optimized geometries. No imaginary frequencies were found in any case. These calculations were performed in vacuum and later repeated under a polarizable continuous solvation model (SMD) with water ($\varepsilon$ = 78.35 D) and *n*-hexane ($\varepsilon$ = 1.88 D) to study their behavior under drastically different polarities; this continuous solvation model is used even at the cost of losing information about microsolvation. Interaction energies between adjacent erucamide molecules were computed under the NBODel method with the NBO3.1 program[22], as provided in G09. This procedure deletes (i.e. sets to zero) all elements of the Fock Matrix which correspond to integrals centered on atoms that belong to different chains. The resulting matrix is diagonalized again with a resulting increase in the total energy and thus the difference is ascribed to the deleted interaction[23]. The same procedure described above was followed for behenamide.

We start by discussing the results for brushes in vacuum. After optimization in vacuum of the pendant molecules, erucamide chains remain closer to each other than in the case of behenamide, where a more spread out configuration is observed (see Fig. 2). Interaction energy values for the middle adjacent erucamide and behenamide chains are 42.05 and 33.85 kcal/mol, respectively (in vacuum model). The reason behind this large energy difference is found to be the interaction between π electrons in the former giving rise to π – π stacking interactions.

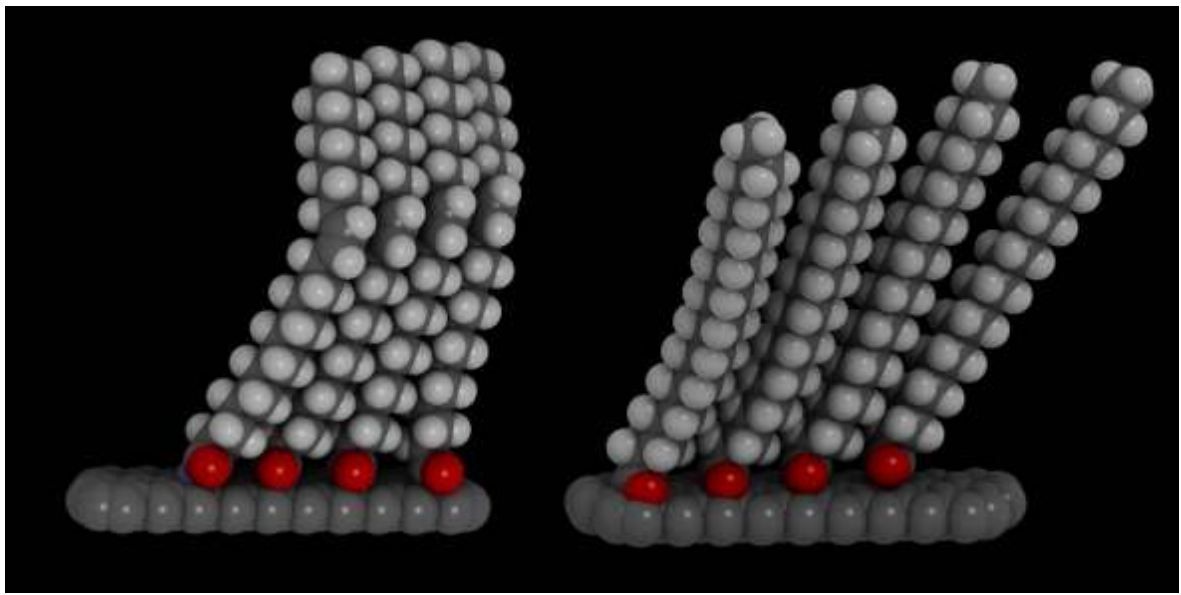

**Figure 2**. Optimized erucamide (*left*) and behenamide (*right*) chains on a 2x10 sheet of zigzag graphene, which was kept frozen throughout all calculations. (Image created with QuteMol[24]).

To confirm this hypothesis a perturbation theory calculation at second order was performed to measure the delocalization energies between chains at these particular atoms. $\pi_i - \pi_j^*$ delocalizations are obtained in the range of 1.21 to 2.17 kcal/mol, which indicates that despite their low interaction values this is a highly organized and cooperative driving force in keeping erucamide molecules bound, which is obviously lost in behenamide due to the absence of these double bonds. The highest occupied molecular orbital (HOMO) for this compound is comprised of the interacting π bonds on all four erucamide molecules, see Fig. 3.

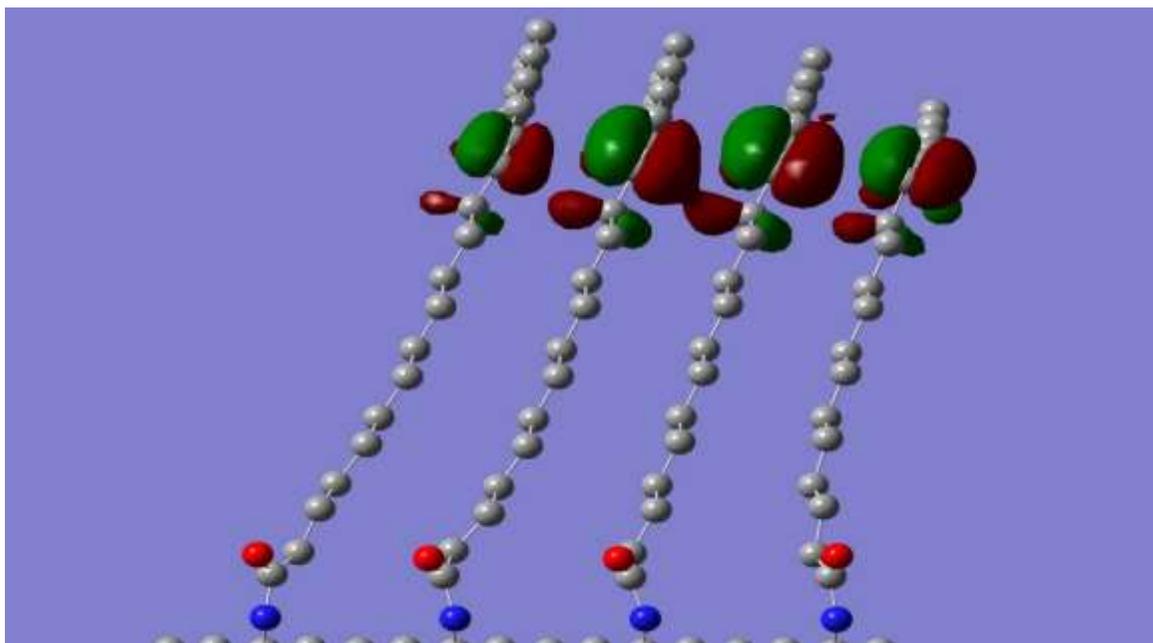

**Figure 3**. Frontal view of the highest occupied molecular orbital (HOMO) for the four interacting erucamide molecules; the π bonds clearly interact between them to increase the overall stability of the brush. Hydrogen bonds and graphene scaffold are omitted for clarity.

Although the surface coverage is the same for both types of chains shown in Fig. 2, the fact that the behenamide chains are more spread out than those that make up the erucamide brush means the solvent penetrates further into the behenamide brush. This in turn means that the COF is larger in such a case, because there is less lubricant left free between the surfaces to reduce friction, as has been found in other reports[18, 19]. Consequently, erucamide brushes reduce friction more than their behenamide counterparts at a given surface coverage.

A mapping of the electrostatic potential onto the electronic isodensity surface of the erucamide and behenamide molecules ($\rho$ = 0.002 e Bohr$^{-3}$) shows that the electrostatic contribution to their interaction is negligible as can be observed in Fig. 4. In it, the coloring indicates that a homogeneous electrostatic charge is spread out over the chains with seemingly no change in the $\pi$ bonds region. A similar potential is observed around C13-C14 for both systems, indicating that the interaction between chains lacks a significant contribution of the electrostatic component; both the highest and lowest potential values are found at the amide group on the bound end of the chains.

Following the optimizations in vacuum both systems were further optimized under an implicit solvent model (SMD) using water and *n*-hexane without much change in their respective geometries. However, the interaction energies between inner chains rise slightly to the following values: (a) water: 44.21; 34.46 kcal/mol and (b) *n*-hexane: 42.11; 33.89 kcal/mol for erucamide and behenamide respectively, which is indicative of a small but detectable solvent influence on the interaction between the chains.

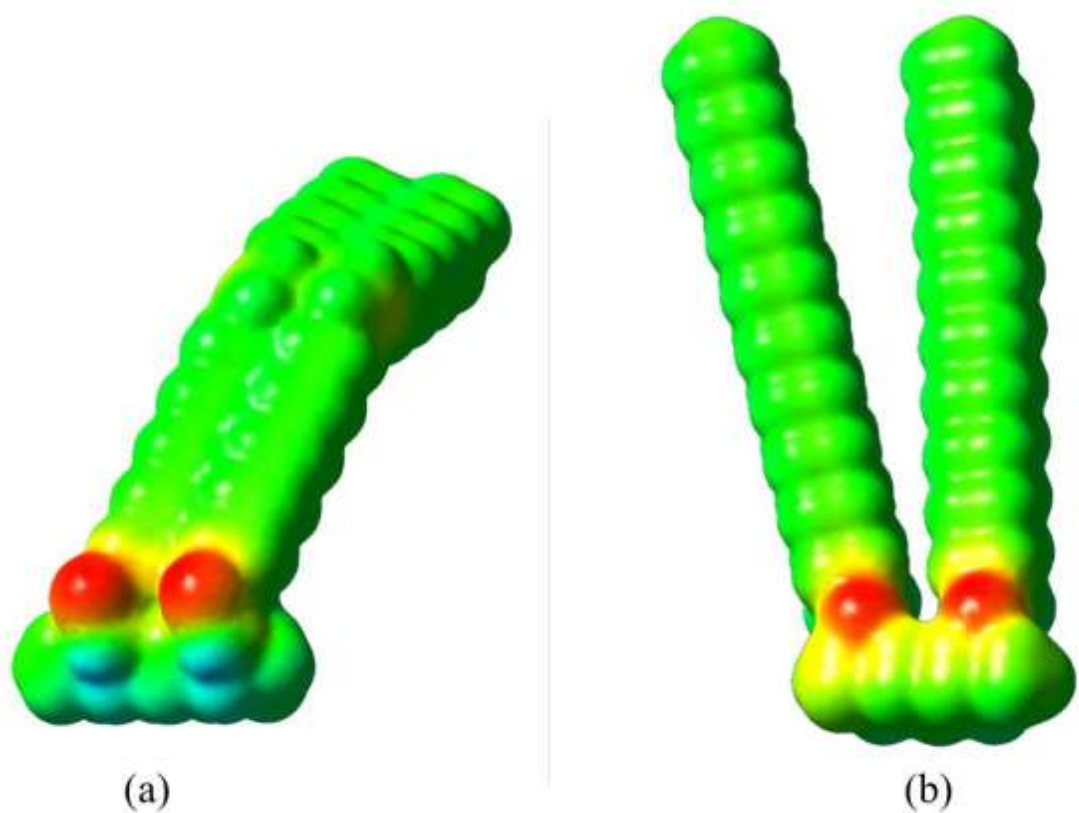

**Figure 4**. Electrostatic potential mapped onto the isodensity electronic surface ($\rho = 0.002$ e Bohr$^{-3}$) for the erucamide (a) and behenamide (b) brushes. Only the central two molecules are shown for the sake of clarity.

As an extension of these calculations a 2D substitution pattern was considered. Seven hydrocarbon chains were grafted on two rows of a larger graphene sheet and the vacuum calculations described above were carried out. As before, each chain was separated from the nearest by four carbon – carbon bonds. Optimized structures for both systems are displayed in Fig. 5. As in the case of the 1D substitution pattern, erucamide chains pile together through π – π stacking interactions, however the increased steric hindrance caused around the cis- double bond conformation yields a more disorganized network. In the case of behenamide the opposite seems to occur, for there seems to be some steric interlocking between the chains leading to a more structured arrangement which in turn is corroborated with the calculation of the interaction energies of the middle chains. The corresponding $E_{int}$ values are 38.97 and 41.60 kcal/mol, respectively, which is the opposite trend to the one observed for the 1D alignment.

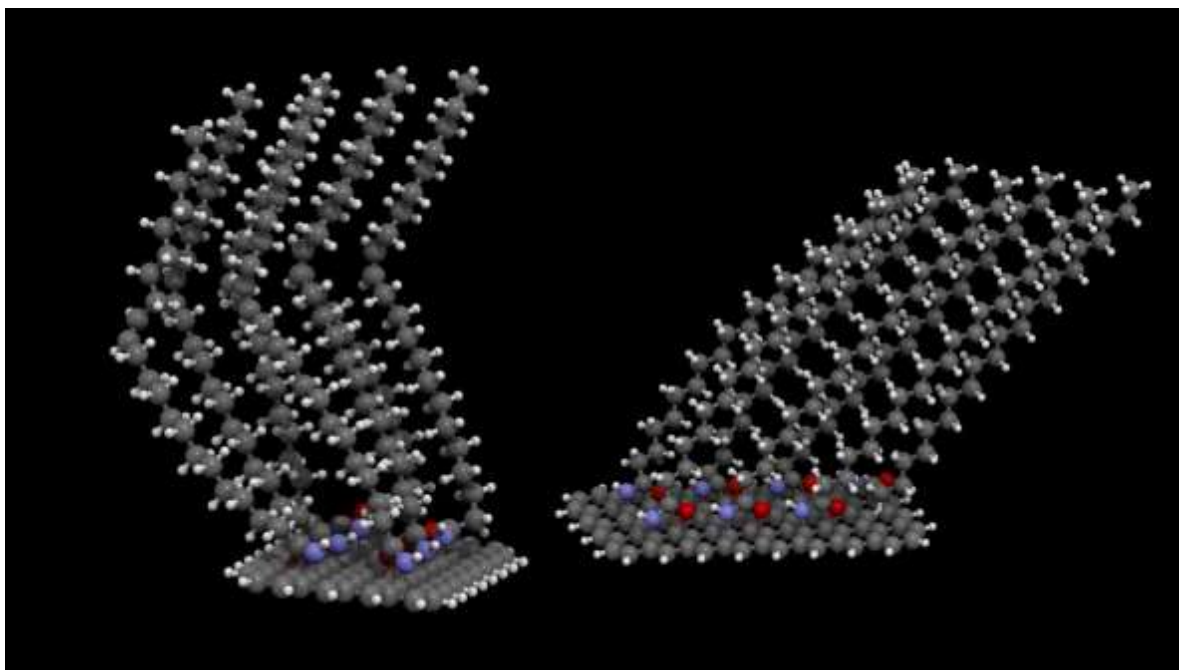

**Figure 5.** Optimized erucamide (*left*) and behenamide (*right*) chains on a 2D substitution pattern on graphene, which was kept frozen throughout all calculations. (Image created with QuteMol[24]).

In conclusion, the addition of slip agents to surfaces to reduce friction between them has become common place in the polymer and coatings industries, to name a few. Two leading agents used for such purpose are the erucamide and behenamide molecules, whose chemical compositions are almost identical, with the only difference being the presence of a double bond between carbon atoms in the erucamide chain. However, it is known from tribology experiments that, under equal conditions, brushes made up of erucamide chains display a COF that can be up to a factor of three smaller than the COF obtained when using behenamide as a slip agent. To trace the origin of such behavior we performed DFT calculations of brushes made up of four chains of each type of amide grafted to a graphene surface, finding that there is a weak but measurable interaction between erucamide chains, which keeps them closer to each other than in the behenamide brush. The presence of $\pi - \pi$ stacking interactions in adjacent erucamide molecules gives rise to a cooperative interaction that keeps these molecules closer together as opposed to the case of behenamide in which such interactions are absent and hence a more spread configuration is observed upon optimization due to a change in energy of ca. 10 kcal/mol. Although the calculations for the two – dimensional arrangement show the erucamide chains disorganize somewhat more than in the one dimensional case, they are still more compactly organized than the behenamide brush, as Fig. 5 illustrates. This many – body organization of the erucamide chains generates a brush that is not easily penetrated by the solvent, which remains at the interface between brushes, playing a lubrication role. Therefore, cooperative quantum mechanical interactions, even if they are weak such as those between adjacent erucamide chains, can lead to markedly different mesoscopic and even macroscopically measurable properties, such as the friction coefficient between surfaces covered with amides. This knowledge is expected to be useful for the design of new soft matter systems, as well as for the basic understanding of the fundamental mechanisms that optimize tribological properties of polymer chains.

## AKCNOWLEDGEMENTS

We thank DGCTIC – UNAM for access granted to the supercomputer known as '*Miztli*', and to Citlali Martínez for keeping our local computing facilities running in optimal conditions. AGG would like to thank M. A. Balderas Altamirano, R. López Esparza, E.

Pérez, Z. Quiñones, E. Rivera and M. A. Waldo for fruitful discussions. The support of Proinnova – CONACYT, through grant 231810 is fully acknowledged.

# REFERENCES

[1] R. M. Overney, E. Meyer, J. Frommer, and H.-J.Güntherodt, "Force Microscopy Study of Friction and Elastic Compliance of Phase-Separated Organic Thin Films", *Langmuir* **10** (1994) 1281-1286.

[2] K. J. Ryan, K. E. Lupton, P. G. Pape, and V. B. John, "UItra-High-Molecular-Weight Functional Siloxane Additives in Polymers. Effects on Processing and Properties", *Journal of Vinyl & Additive Technology*, **6** (2000) 7-19.

[3] A. S. Rawls and D. E. Hirt, "Evaluation of Surface Concentration of Erucamide in LLDPE Films", *Journal of Vinyl & Additive Technology*, **8** (2002) 130-138.

[4] M. Mansha, C. Gauthier, P. Gerard, and R. Schirrer, "The effect of plasticization by fatty acid amides on the scratch resistance of PMMA", *WEAR*, **271** (2011) 671-679.

[5] C. A. Shuler, A. V. Janorkar, and D. E. Hirt, "Fate of Erucamide in Polyolefin Films at Elevated Temperature", *Polymer Engineering and Science*, **44** (2004) 2247-2253.

[6] F. Coelho, L. F. Vieira, R. Benavides, M. Marques da Silva Paula, A. M. Bernardin, R. F. Magnago, and L. da Silva, "Synthesis and Evaluation of Amides as Slip Additives in Polypropylene", *International Polymer Processing*, **30** (2015) 574-584.

[7] M. X. Ramírez, D. E. Hirt, and L. L. Wright, "AFM Characterization of Surface Segregated Erucamide and Behenamide in Linear Low Density Polyethylene Film", *Nano Letters*, **2** (2002) 9-12.

[8] L. Walp and H. Tomlinson, "Effect of Erucic Acid Percentage on Slip Properties of Erucamide", *Journal of Plastic Film & Sheeting*, **20** (2004) 275-287.

[9] M. X. Ramírez, K. B. Walters, and D. E. Hirt, "Relationship between Erucamide Surface Concentration and Coefficient of Friction of LLDPE Film", *Journal of Vinyl & Additive Technology*, **11** (2005) 9-12.

[10] C. M. Allen and E. Drauglis, "Boundary Layer Lubrication: Monolayer or Multilayer", *WEAR*, **14** (1969) 363-384.

[11] P. U. Dhanvijay, V. D. Gharat, and V. V. Shertukde, "Synthesis and characterization of slip additive functioning as an intercalating agent", *International Journal of Plastics Technology*, **18** (2014) 100-112.

[12] C. L. Swanson, D. A. Brug, and R. Kleiman, "Meadowfoam Monoenoic Fatty Acid Amides as Slip and Antiblock Agent in Polyolefin Film", *Journal of Applied Polymer Science*, **49** (1993) 1619-1624.

[13] E. Har-Even, A. Brown, and E. I. Meletis, "Effect of friction on the microstructure of compacted solid additive blends for polymers", *WEAR*, **328-329** (2015) 160-166.

[14] S. Loehle, "Understanding of adsorption mechanisms and tribological behaviors of C18 fatty acids on iron-based surfaces: A molecular simulation approach", Ph.D. thesis, Universite de Lyon (2014).

[15] S. Campen, J. Green, G. Lamb, D. Atkinson, and H. Spikes, "On the Increase in Boundary Friction with Sliding Speed", *Tribology Letters*, **48** (2012) 237-248.

[16] S. Loehle, C. Matta, C. Minfray, T. Le Mogne, J-M. Martin, R. Iovine, Y. Obara, R. Miura, and A. Miyamoto, "Mixed Lubrication with C18 Fatty Acids: Effect of Unsaturation", *Tribology Letters*, **53** (2014) 319-328.

[17] J. P. Ewen, C. Gattinoni, N. M. Morgan, H. Spikes, and D. Dini, "Non-equilibrium molecular dynamics simulations of organic friction modifiers adsorbed on iron oxide surfaces", *Langmuir*, **32** (2016).

[18] M. Doig, C. P. Warrens, and P. J. Camp. "Structure and Friction of Steric Acid and Oleic Acid Films Adsorbed on Iron Oxide Surfaces in Squalane", *Langmuir*, **30** (2014) 186-195.

[19] A. Gama Goicochea, R. López-Esparza, M. A. Balderas Altamirano, E. Rivera-Paz, M. A. Waldo, E. Pérez. "Friction coefficient and viscosity of polymer brushes with and without free polymers as slip agents", *J. Mol. Liq*. **219** (2016) 368-376.

[20] Gaussian 09, Revision D.01, M. J. Frisch, G. W. Trucks, H. B. Schlegel, G. E. Scuseria, M. A. Robb, J. R. Cheeseman, G. Scalmani, V. Barone, B. Mennucci, G. A. Petersson, H. Nakatsuji, M. Caricato, X. Li, H. P. Hratchian, A. F. Izmaylov, J. Bloino, G. Zheng, J. L. Sonnenberg, M. Hada, M. Ehara, K. Toyota, R. Fukuda, J. Hasegawa, M. Ishida, T. Nakajima, Y. Honda, O. Kitao, H. Nakai, T. Vreven, J. A. Montgomery, Jr., J. E. Peralta, F. Ogliaro, M. Bearpark, J. J. Heyd, E. Brothers, K. N. Kudin, V. N. Staroverov, R. Kobayashi, J. Normand, K. Raghavachari, A. Rendell, J. C. Burant, S. S. Iyengar, J. Tomasi, M. Cossi, N. Rega, J. M. Millam, M. Klene, J. E. Knox, J. B. Cross, V. Bakken, C. Adamo, J. Jaramillo, R. Gomperts, R. E. Stratmann, O. Yazyev, A. J. Austin, R. Cammi, C. Pomelli, J. W. Ochterski, R. L. Martin, K. Morokuma, V. G. Zakrzewski, G. A. Voth, P. Salvador, J. J. Dannenberg, S. Dapprich, A. D. Daniels, Ö. Farkas, J. B. Foresman, J. V. Ortiz, J. Cioslowski, and D. J. Fox, Gaussian, Inc., Wallingford CT, 2009.

[21] S. Grimme, "Semiempirical GGA-type density functional constructed with a long-range dispersion correction", *J. Comp. Chem*. **27** (2006) 1787-99.

[22] NBO Version 3.1, E. D. Glendening, A. E. Reed, J. E. Carpenter, and F. Weinhold.

[23] A. E. Reed, L. A. Curtiss, and F. Weinhold, "Intermolecular interactions from a natural bond orbital, donor-acceptor viewpoint", *Chem. Rev*. **88** (1988) 899-926.

[24] Marco Tarini, Paolo Cignoni, Claudio Montani "Ambient Occlusion and Edge Cueing for Enhancing Real Time Molecular Visualization" *IEEE Transactions on Visualization and Computer Graphics* Volume 12 , Issue 5 , Pages 1237-1244 , **2006** , ISSN:1077-2626.